\documentclass[aps,prb,twocolumn,amsmath,amssymb,showpacs,superscriptaddress]{revtex4-1}

\usepackage{graphicx}
\usepackage{amsmath,amssymb,amsfonts}
\usepackage{textcomp}
\usepackage{hyperref}
\usepackage{gensymb}
\usepackage{verbatim}
\usepackage{amsmath}
\hypersetup{breaklinks=true,colorlinks=true,urlcolor=black}
\usepackage{color}
\usepackage{soul}
\DeclareGraphicsExtensions{.jpg,.pdf,.png,.eps}
\usepackage{pdfpages}

\makeatletter

\AtBeginDocument{\let\LS@rot\@undefined}

\makeatother

\begin{document}

\title{Exceedingly Small Moment Itinerant Ferromagnetism of Single Crystalline La$_{5}$Co$_{2}$Ge$_{3}$}
\author{S. M. Saunders}  
\affiliation{Ames Laboratory, U.S. DOE, Iowa State University, Ames, Iowa 50011, USA}
\affiliation{Department of Physics and Astronomy, Iowa State University, Ames, Iowa 50011, USA}
\author{L. Xiang}  
\affiliation{Ames Laboratory, U.S. DOE, Iowa State University, Ames, Iowa 50011, USA}
\affiliation{Department of Physics and Astronomy, Iowa State University, Ames, Iowa 50011, USA}
\author{R. Khasanov}  
\affiliation{Laboratory for Muon Spin Spectroscopy, Paul Scherrer Institute, 5232 Villigen, Switzerland}
\author{T. Kong}  
\affiliation{Ames Laboratory, U.S. DOE, Iowa State University, Ames, Iowa 50011, USA}
\affiliation{Department of Physics and Astronomy, Iowa State University, Ames, Iowa 50011, USA}
\author{Q. Lin}  
\affiliation{Ames Laboratory, U.S. DOE, Iowa State University, Ames, Iowa 50011, USA}
\affiliation{Department of Chemistry, Iowa State University, Ames, Iowa 50011, USA}
\author{S. L. Bud'ko}  
\affiliation{Ames Laboratory, U.S. DOE, Iowa State University, Ames, Iowa 50011, USA}
\affiliation{Department of Physics and Astronomy, Iowa State University, Ames, Iowa 50011, USA}
\author{P. C. Canfield}
\affiliation{Ames Laboratory, U.S. DOE, Iowa State University, Ames, Iowa 50011, USA}
\affiliation{Department of Physics and Astronomy, Iowa State University, Ames, Iowa 50011, USA}

\begin{abstract}
Single crystals of monoclinic La$_{5}$Co$_{2}$Ge$_{3}$ were grown using a self-flux method and were characterized by room-temperature powder X-ray diffraction, anisotropic temperature and field dependent magnetization, temperature dependent resistivity, specific heat, and muon spin rotation. La$_{5}$Co$_{2}$Ge$_{3}$ has a Curie temperature ($T_\mathrm{C}$) of 3.8~K and clear signatures of ferromagnetism in magnetization and $\mu SR$ data, as well as a clear loss of spin disorder scattering in resistivity data and a sharp specific heat anomaly. The magnetism associated with La$_{5}$Co$_{2}$Ge$_{3}$ is itinerant, has a change in the entropy at $T_\mathrm {C}$ of $\simeq$0.05 R ln2 per mol-Co, and has a low-field saturated moment of $\sim 0.1 \mu_\mathrm B$/Co, making it a rare, itinerant, small moment, low $T_\mathrm C$ compound.
\end{abstract}

\maketitle

 
Magnetism in metallic compounds has typically been described in a local moment or itinerant moment picture. The local moment description has been studied across many systems, due, in part, to the convenience of rare-earth elements containing partially filled 4-f shells which provide well defined, local magnetic moments\cite{Szytula_RE_IM}. There are fewer examples of itinerant magnetism, especially ferromagnetic systems with very low $T_\mathrm C$ and $\mu_{sat}$. For example,  Sc$_3$In~\cite{Matthias_1961}, ZrZn$_2$\cite{Matthias_1958}, MnSi\cite{Kadowaki1982}, LuFe$_2$Ge$_2$\cite{Avila2004, Fujiwara2007} and TiAu\cite{Svanidze_2015} have been suggested to be itinerant with low transition temperatures: $T_\mathrm C = 6$ K and 35 K for the ferromagnetic Sc$_3$In and ZrZn$_2$, respectively, and $T_\mathrm N = 29$ K, 9 K and 36 K for the antiferromagnetic MnSi, LuFe$_2$Ge$_2$ and TiAu, respectively.

In this letter, we report the discovery and basic properties of the itinerant ferromagnet (IFM) La$_5$Co$_2$Ge$_3$. La$_5$Co$_2$Ge$_3$ is composed of 50\% non-moment-bearing La,  30\% Ge, and only 20\% Co; transport and thermodynamic measurements exhibit a Curie temperature of $T_\mathrm C = (3.8\pm0.1)~K$, which is one of the lowest reported transition temperatures for an ordered, stoichiometric IFM. Temperature and field dependent magnetization measurements reveal $\mu_{eff} = (1.10\pm0.05)~\mu_{B}$/Co whereas the low-field $\mu_{sat} = 0.1~\mu_{B}$/Co leading to a Rhodes-Wohlfarth ratio~\cite{RhodesWohlfarth1963} of 4.9. In addition, specific heat data show a greatly reduced loss of entropy, 0.05 R ln2 per mol-Co, associated with the transition. Muon spin rotation ($\mu$SR) measurements indicate static moments and internal fields consistent with a greatly reduced ordered moment magnitude when compared to full-moment Co. 
  
\begin{figure}[tbh!]
\includegraphics[scale=0.3]{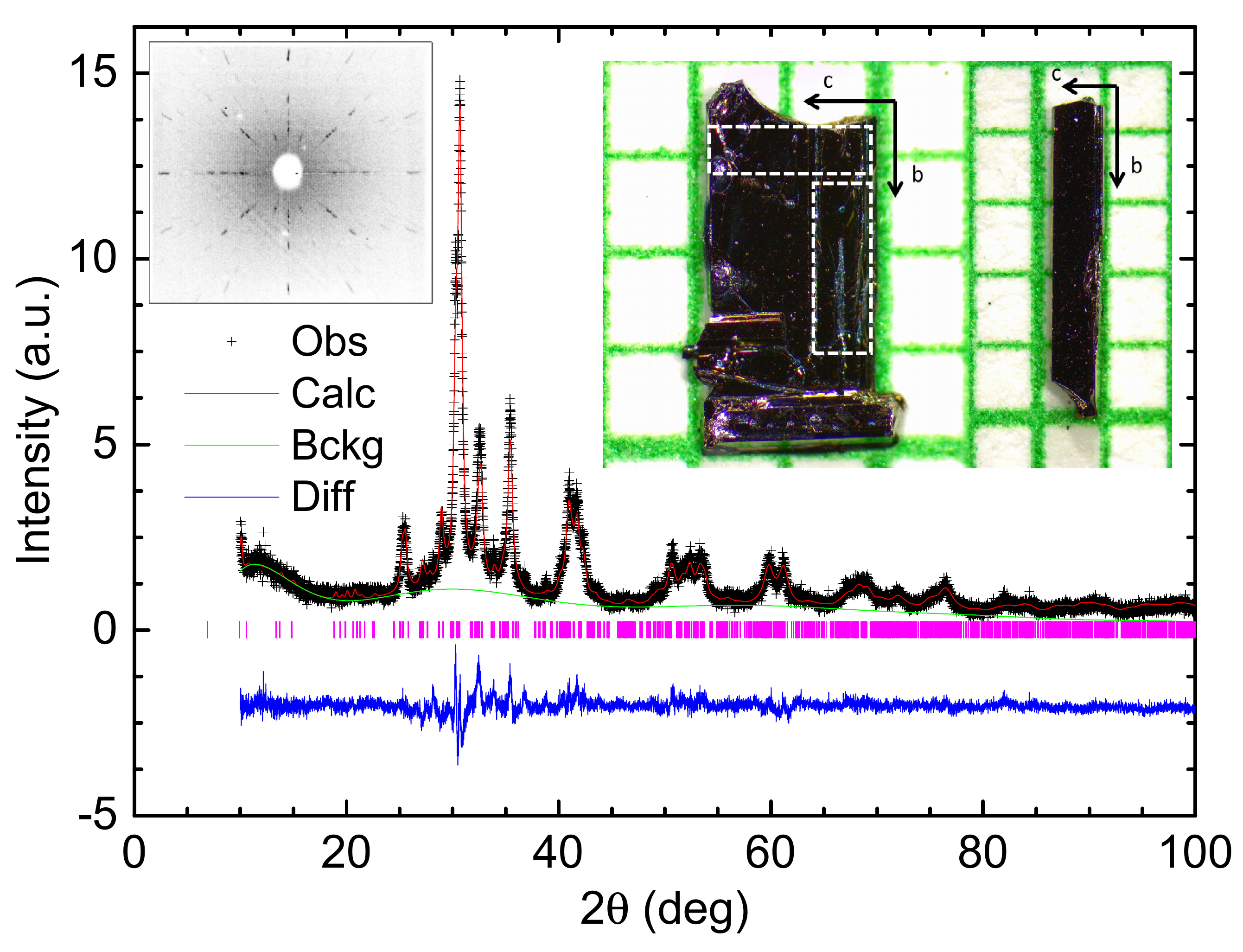}  
\caption{Powder x-ray diffraction data for La$_{5}$Co$_{2}$Ge$_{3}$. Vertical pink lines represent expected peak positions for structural data (Tables S1 and S2 in Supplemental Information). Inset:(left) back reflection Laue diffraction pattern with beam perpendicular to the face of the plate, showing the two-fold mirror symmetry expected for a monoclinic system. (right) Representative crystal morphologies of La$_{5}$Co$_{2}$Ge$_{3}$ shown on mm grid. A typical plate (left) and blade(right) are shown. Dashed lines outline samples which were cut and measured for resistivity (see text).}
\label{fig:xraydata}
\end{figure}

Single crystals of La$_{5}$Co$_{2}$Ge$_{3}$ were grown using a self-flux solution growth method~\cite{Fisher_Canfield_solution_growth, Canfield_progress_2020, Lin_2017}. The initial composition of the three elements was La:Co:Ge = 45:45:10. The starting elements (Co (99.9\%), Ge (Alfa Aesar 99.9+\%), La (Ames Lab 99.9\%)) were combined in a 3-cap tantalum crucible \cite{Fisher_Canfield_solution_growth, Canfield_progress_2020} and sealed in a fused silica ampoule under a partial argon atmosphere. The ampoule was then heated to 1180$^\circ$ C, held at 1180$^\circ$ C for 4 hours and slowly cooled to 800$^\circ$ C over 40 hours at which point the remaining solution was decanted with the assistance of a centrifuge. The crystals of La$_{5}$Co$_{2}$Ge$_{3}$ grew in thin plates as well as long blades, as shown in Fig.~\ref{fig:xraydata}. The crystals are not air sensitive.

La$_{5}$Co$_{2}$Ge$_{3}$ is isostructural to Pr$_{5}$Co$_{2}$Ge$_{3}$~\cite{Lin_2017}; the crystal structure was established at room temperature and ambient pressure using a Rigaku Miniflex powder X-ray diffractometer (Cu Kα radiation). Samples were prepared by grinding a single crystal into powder, which was then mounted and measured on a single crystal Si, zero-background sample holder. A typical X-ray diffraction pattern, where all major peaks are consistent with the La$_{5}$Co$_{2}$Ge$_{3}$ monoclinic structure, is shown in Fig.~\ref{fig:xraydata} and discussed in further detail in the Supplemental Information section. When growing La$_{5}$Co$_{2}$Ge$_{3}$, two morphologies emerged in the growth crucible, with representative examples shown in Fig.~\ref{fig:xraydata}. However, when studied by powder x-ray diffraction, the powder x-ray patterns for plate-like and blade-like crystals are identical. As determined by back reflection Laue diffraction, the direction perpendicular to the face of the crystal is the $a^*$ direction, which is perpendicular to $b$ and $c$.

Back reflection Laue images were collected at room temperature. The incident x-rays were produced by a 40~kV and 15~mA power source through a 0.5~mm diameter circular aperture and collected over 300~s. Crystal systems with a monoclinic unit cell (Fig. S1), like La$_{5}$Co$_{2}$Ge$_{3}$, are part of the $2/m$ Laue class. As such, they will exhibit two-fold symmetry in the back reflection pattern, which is shown in the inset to Fig.~\ref{fig:xraydata}. Using this image and the corresponding unit cell data (Table S1 \cite{SI}), the peaks were indexed with the assistance of CLIP (the Cologne Laue Indexation Program)~\cite{CLIP4} and the specific orientation of the crystal that would give rise to the resultant peaks was identified.


DC magnetization measurements were performed in a Quantum Design Magnetic Property Measurement System 3 (MPMS 3), superconducting quantum interference device (SQUID) magnetometer ($T$ = 1.8 - 300 K, $H_{max}$ = 70 kOe). All samples were manually aligned to measure the magnetization along the desired axis. A blade-like crystal was selected with measurements performed perpendicular to the face of the blade and parallel to the face of the blade. Measurements conducted perpendicular to the blade are perpendicular to the b-c plane (i.e.parallel to a$^*$). Samples which were aligned parallel to the plate are in either the b or c direction (see Fig.~\ref{fig:xraydata}). For measurements with $H\vert\vert$b or c, the sample was mounted on a quartz rod and attached by GE varnish.

Resistivity measurements were performed using a standard four-probe technique with the temperature environment provided by a MPMS with I = 1~mA supplied by an LR-700 resistance bridge. As shown in Fig.~\ref{fig:xraydata}, plate-like samples allowed for the creation of samples that had current along the b-axis or the c-axis. Epotek-H20E epoxy was used to connect Pt wires to the sample so that the current was flowing in the desired direction.

Specific heat measurements between $T$=1.8~K and 50~K were performed in a Quantum Design Physical Property Measurement System (PPMS) utilizing the relaxation technique with fitting of the whole temperature response of the micro-calorimeter. A plate-like sample was mounted on the micro-calorimeter platform using a small amount of the Apiezon N grease. A 2\% temperature rise at each measurement point was used. The addenda (contribution from the grease and sample platform) was measured separately and subtracted from the data using PPMS software.

The Zero-field muon-spin rotation ($\mu$SR) measurements were performed at the $\pi$E1 beamline by using Dolly spectrometer (Paul Scherrer Institute, PSI Villigen, Switzerland). The $^4$He cryostat equipped with the $^3$He inset (base temperature $\simeq 0.26$~K) was used. Samples were mounted on a thin copper foil ($\simeq 10$~$\mu$m), which was transparent for positive surface muons used in our studies. 

\begin{figure}[tbh!]
\includegraphics[scale=0.3]{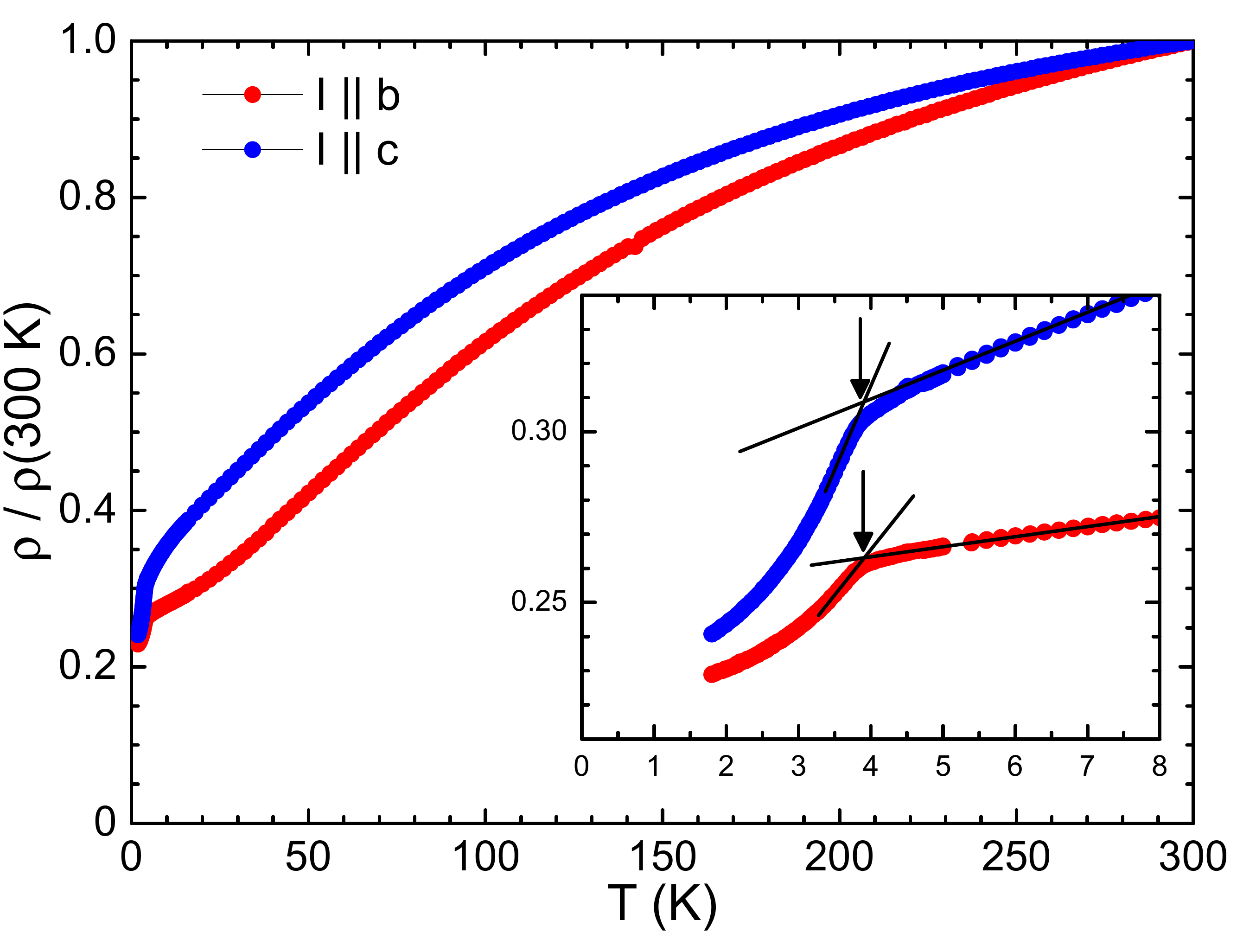} 
\caption{Zero-field, normalized, in-plane resistivity versus temperature for current flowing along b- or c-axis (inset: low-temperature zoom of resistivity versus temperature with the criterion for determining $T_\mathrm C$ indicated by the lines and arrows). At $T = 300$~K, $\rho_b=220 ~\mu\Omega$-cm and $\rho_c=390 ~\mu\Omega$-cm}
\label{fig:resistivity}
\end{figure}

Resistivity measurements (Fig.~\ref{fig:resistivity}) show that the samples are metallic; at $T = 300$~K, $\rho_b=220 ~\mu\Omega$-cm and $\rho_c=390 ~\mu\Omega$-cm. The crystals that were measured have residual resistance ratios (RRR=$\rho$(300~K)/$\rho$(2~K)) ranging from 3 - 5. Below $T = 4$~K there is a sharp drop in resistivity with an onset temperature, $T_\mathrm C = 3.8$ K, indicated by the arrows in the inset of Fig.~\ref{fig:resistivity}.

\begin{figure}[tbh!]
\includegraphics[scale=0.3]{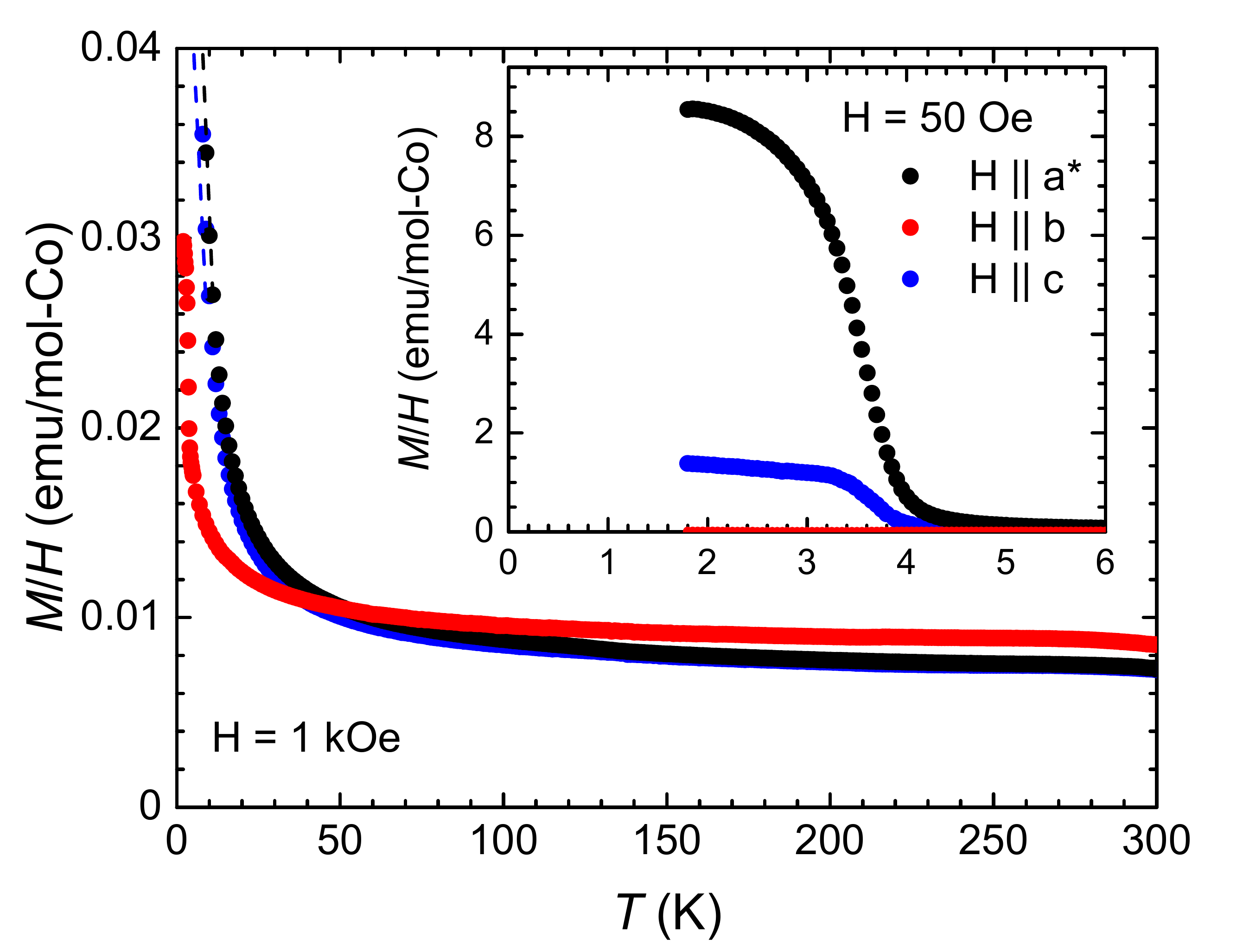} 
\caption{Anisotropic magnetic susceptibility of La$_{5}$Co$_{2}$Ge$_{3}$ measured at $H$ = 1~kOe (inset: low-temperature zoom of anisotropic magnetic susceptibility at $H$ = 50~Oe).}
\label{fig:mt_perco}
\end{figure}

The temperature dependent magnetization, $M(T)$, for La$_{5}$Co$_{2}$Ge$_{3}$ is shown in Fig.~\ref{fig:mt_perco}. The low temperature, $H$ = 50~Oe, $M(T)$ data show a clear transition below 4.0~K. The higher temperature, $H$ = 1~kOe, $M(T)$ data manifest a clear Curie-Weiss-like behavior for 20~K$\leq T\leq$ 100~K. When the data are fit to,
\begin{equation} \label{eqn:CWwithChi0}
\frac{M}{H} = \frac{C}{T - \Theta} + \chi_0
\end{equation}
with Curie Constant C$=N(\mu_{eff}\mu_{B})^2/3k_B$, values of $\mu_{eff} = 1.17 ~\mu_{B}/Co$, $\Theta = 0.5$ K, $\chi_0 = 0.007$, $\mu_{eff} = 1.03 ~\mu_{B}/Co$, $\Theta = -13$ K, $\chi_0 = 0.008$, and $\mu_{eff} = 1.08 ~\mu_{B}/Co$ $\Theta = 1.3$ K $\chi_0 = 0.007$ were found for $H$ parallel to the $a^*$, $b$, and $c$ directions, respectively. For all directions, uncertainties for $\mu_{eff}$ and $\Theta$ are determined to be  $\pm0.1~\mu_{B}/$Co and $\pm4$ K respectively, due primarily to the uncertainties in the measurement of the mass. $M(T)$ data collected on a significantly larger, polycrystalline sample (see Supplemental Information Fig. S2) gave values of $\mu_{eff} = (1.10\pm0.05) ~\mu_{B}/Co$ and $\Theta = (-10.7\pm0.2)$ K from a fit for 10~K$\leq T \leq$ 300~K.

\begin{figure}[tbh!]
	\includegraphics[scale=0.3]{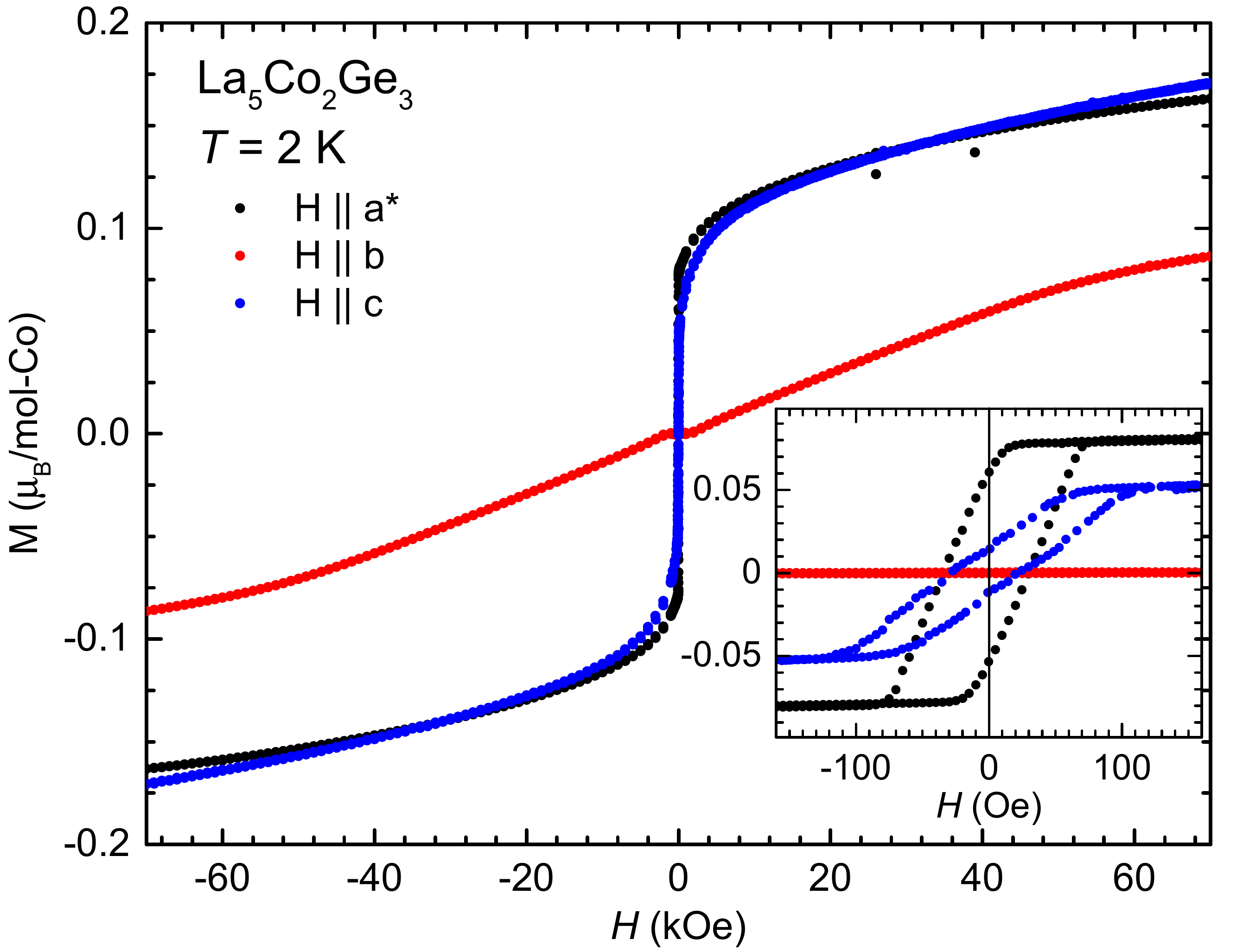} 
	\caption{Anisotropic magnetization versus field isotherms of La$_{5}$Co$_{2}$Ge$_{3}$. (Inset: low field zoom of data revealing hysteresis for $H ~\vert \vert~ a^* $ and $H ~\vert \vert~c$.)}
	\label{fig:La_MH_toShow}
\end{figure}

Anisotropic magnetization versus field data (Fig. \ref{fig:La_MH_toShow}) were taken for $\vert H\vert \leq$ 70~kOe at $T$ = 2~K. A striking anisotropy is readily apparent. Whereas for $H\vert\vert a^*$ and $H\vert\vert c$ there is a low-field saturation to an $\simeq$0.1 $\mu_{B}$ per mol-Co value, for $H\vert\vert b$ the $M(H)$ data has no such feature. For fields well above their initial saturations, the $H\vert\vert a^*$ and $H\vert\vert c$ $M(H)$ data show a very similar, gradual increase with $H$ as does the $H\vert\vert b$ data. The inset to Fig.~\ref{fig:La_MH_toShow} shows that for the two easier axes there is clear hysteresis that can be associated with domain pinning.  Utilizing a linear fit of the data just above saturation to extrapolate to $H$ = 0, we obtain $\mu_{sat} = 0.08~\mu_B$ per mol Co, $\mu_{sat} = 0~\mu_B$ per mol Co, and $\mu_{sat} = 0.05~\mu_B$ per mol Co for the $a^*$, $b$, and $c$ directions, respectively.

Taken as a whole, the $M(T,H)$ data shown in Figures \ref{fig:mt_perco} and \ref{fig:La_MH_toShow} suggest that below $T_\mathrm C$, La$_{5}$Co$_{2}$Ge$_{3}$ becomes a small moment, easy-plane ferromagnet that has a more isotropic, non-linear, but smoothly varying $M(H)$ behavior superimposed on top of the low field saturation.

\begin{figure}[tbh!]
	\includegraphics[scale=0.3]{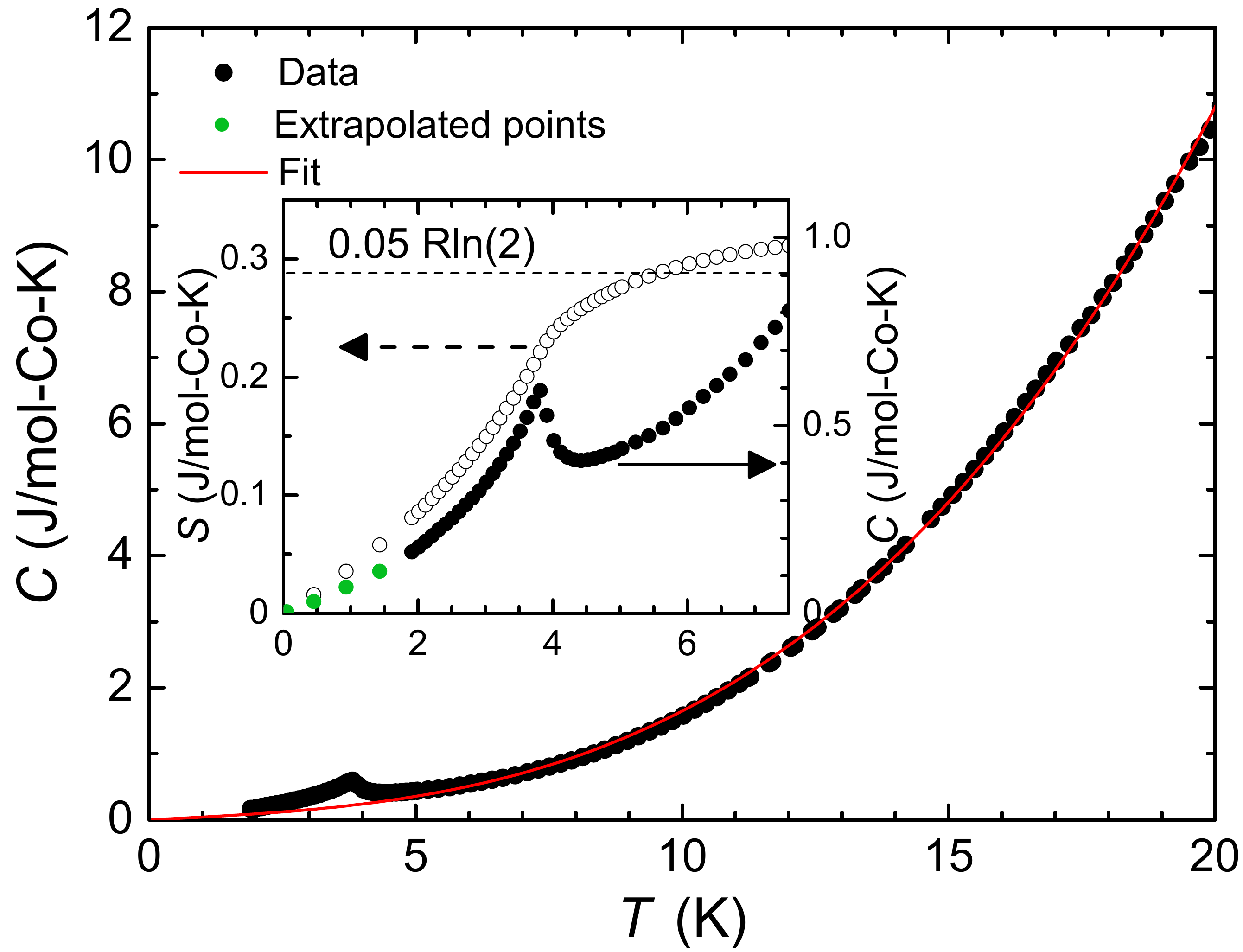} 
	\caption{Heat capacity versus temperature (inset: low temperature zoom of entropy versus temperature (left axis) and heat capacity versus temperature (right axis) data was extrapolated to 0, 0 (green points) to allow for evaluation of entropy). Red line shows Debye fit to data (see text).}
	\label{fig:specificheat}
\end{figure}

Specific heat data, as shown in Fig.~\ref{fig:specificheat}, exhibit a cusp with a maxima at  $T$ = 3.8~K. Given that our resistivity data show a similar transition at 3.8~K and our low field $M(T)$ data show a sharp rise around  3.9~K, we conclude that La$_5$Co$_2$Ge$_3$ becomes ferromagnetic below $T_\mathrm C$ = (3.8 $\pm$ 0.1)~K.

Specific heat data were fit using $C = \gamma T + \beta T^3$ over the region 10~K $<$ $T$ $<$ 15~K which is linear in C/T vs T$^2$. Through this fit, we obtain coefficients of $\gamma \approx 40$~mJ/mol-K$^2$ and $\beta \approx 2.7$~mJ/mol-K$^4$. We then used these fitted values of $\gamma$ and $\beta$ to extrapolate data points to $T$ = 0~K and to estimate the electron and phonon contributions to the specific heat. To estimate the entropy associated with the magnetic transition, we subtracted the inferred electron and phonon contributions from the specific heat data and integrated with respect to $T$. The entropy inferred from the specific heat data (inset of Fig. \ref{fig:specificheat}) reveals the total magnetic entropy of the transition is roughly 0.05 R ln(2) per Co.

\begin{figure}[tbh!]
\includegraphics[scale=0.5]{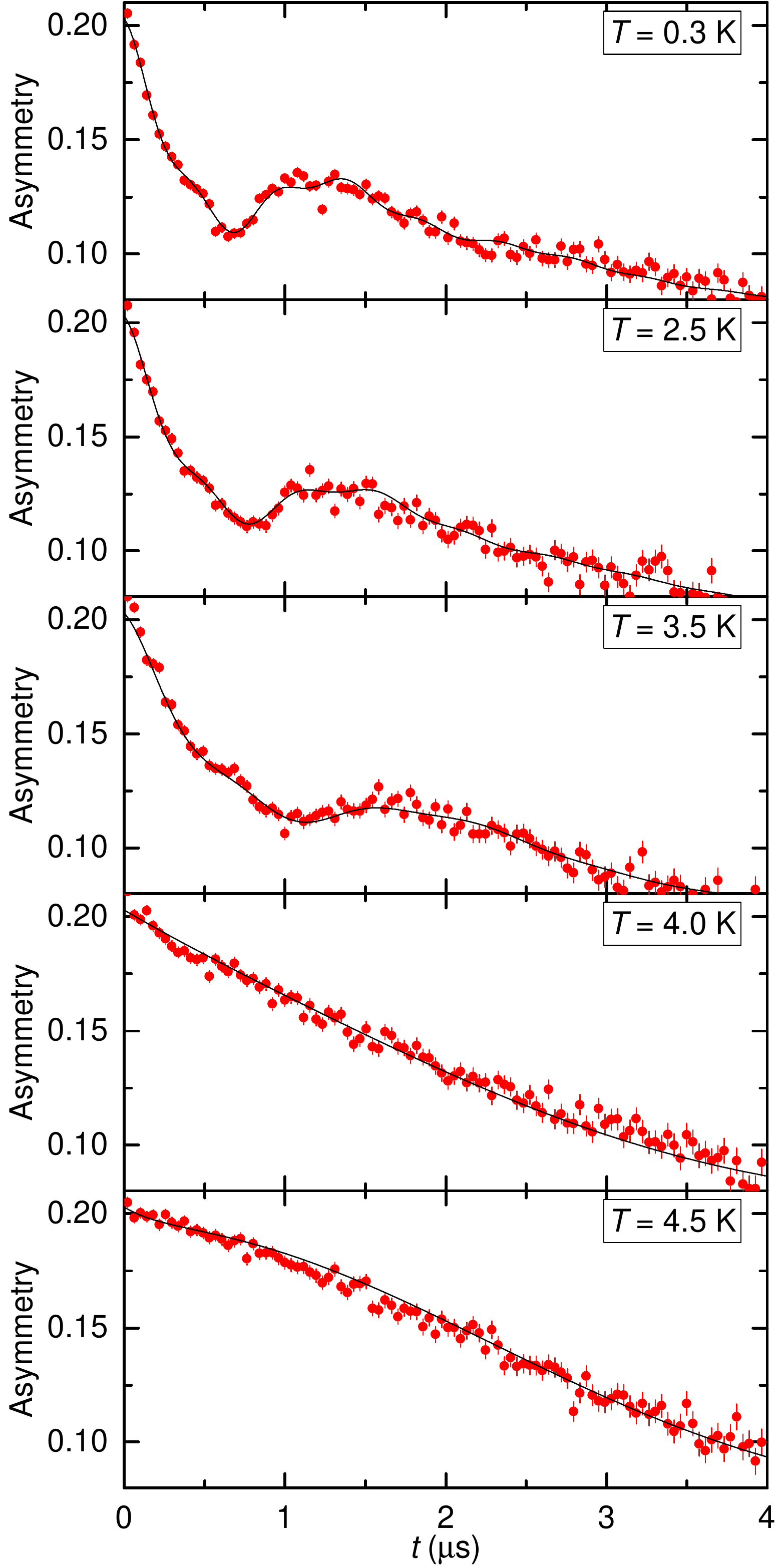} 
\caption{Zero-field $\mu$SR spectra data of La$_{5}$Co$_{2}$Ge$_{3}$. Solid lines are fit made with two cosine signals with zero initial phase.}
\label{zero_field_musr}
\end{figure}

\begin{figure}[tbh!]
\includegraphics[scale=0.3]{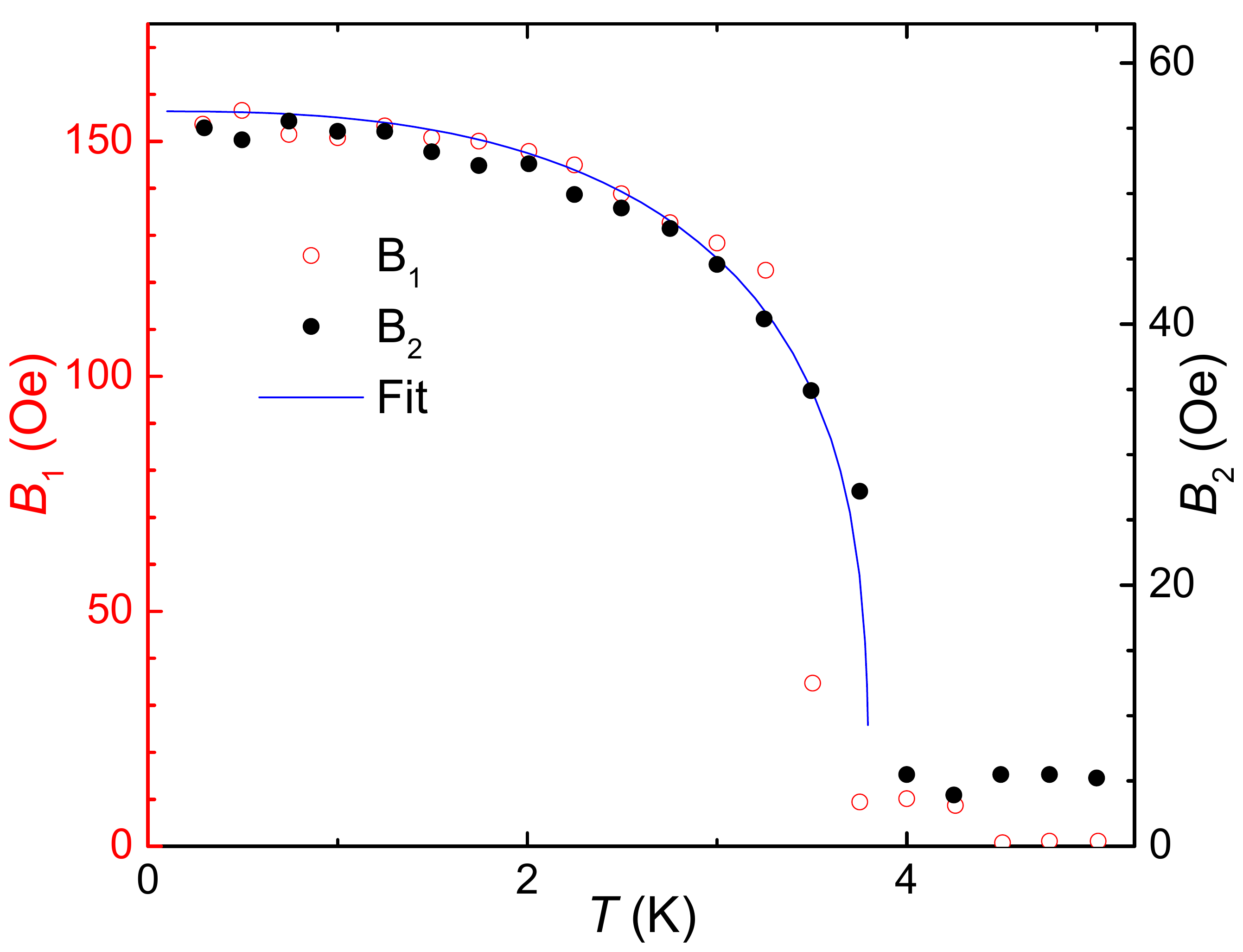}
\caption{Internal field versus Temperature of La$_{5}$Co$_{2}$Ge$_{3}$. The fit was made by assuming two independent oscillatory components. The internal fields (B$_1$ and B$_2$) and the transversal relaxations ($\Lambda_1$ and $\Lambda_2$) were assumed to be different. The longitudinal relaxations, $\Lambda_L$, were assumed to be the same. The relative volume fraction corresponding to the higher field is $\sim$20\% and the one corresponding to the lower field is $\sim$80\%.}
\label{musr_intfield}
\end{figure}

Taken together, the data so far strongly suggest that La$_{5}$Co$_{2}$Ge$_{3}$ is a small moment ferromagnet; in order to test this microscopically, we performed $\mu$SR measurements on a sample for 0.26~K $<$ $T$ $<$ 5~K 
(Figs.~\ref{zero_field_musr}, \ref{musr_intfield} and S3). For $T$ $<$ 4~K we found static magnetic order. The magnetic order is found to be commensurate, as the fit in Fig.~\ref{zero_field_musr} was made with two cosine signals with zero initial phase. The presence of two internal fields suggests the presence of two sites within the crystal lattice where the muons come into rest. $\sim$20\% of muons stop in a higher field site and $\sim$80\% stop in a lower field site (Fig.~\ref{musr_intfield}).

Although La$_{5}$Co$_{2}$Ge$_{3}$ does order magnetically, we do not observe internal fields consistent with full Co moments (see Fig.~\ref{musr_intfield}). The larger internal field, $B_{int,1}$, only reaches 150~Oe, which is approximately one order of magnitude smaller than expected for full Co moments~\cite{Jongh1974}, a result consistent with our small saturated moment (Fig.~\ref{fig:La_MH_toShow}) and $\Delta$S $\simeq$ 0.05 R ln(2) per Co (Fig.~\ref{fig:specificheat}). Both of the internal fields exhibit similar temperature dependencies; when fitting the data in Fig.~\ref{musr_intfield} to the power law $B=B_0(1-(T/T_\mathrm C)^{\alpha})^\beta$, we find that $\beta$=0.293 which is consistent with three-dimensional magnetic order ($\beta_{3D}$=1/3).

The difference between the effective moment inferred from magnetization versus temperature data (Fig.~\ref{fig:mt_perco}) and the low-field saturated moment from magnetization versus field data (Fig.~\ref{fig:La_MH_toShow}) can be understood by considering the Rhodes-Wohlfarth ratio $q_c/q_s$,~\cite{Matthias_1961,RhodesWohlfarth1963, Santiago_2017}, where,
\begin{eqnarray}
\lefteqn{\mu_{eff}^2 = q_c(q_c + 2)\mu_B^2} \nonumber \\
&& \mu_{sat} = q_s \mu_B \Rightarrow \nonumber \\ 
&& q_c/q_s = \left(-1 + \sqrt{1 + (\mu_{eff}/\mu_B)^2}\right)/(\mu_{sat}/\mu_B)
\label{eqn:qcONqs}
\end{eqnarray}

\begin{figure}[tbh!]
\includegraphics[scale=0.3]{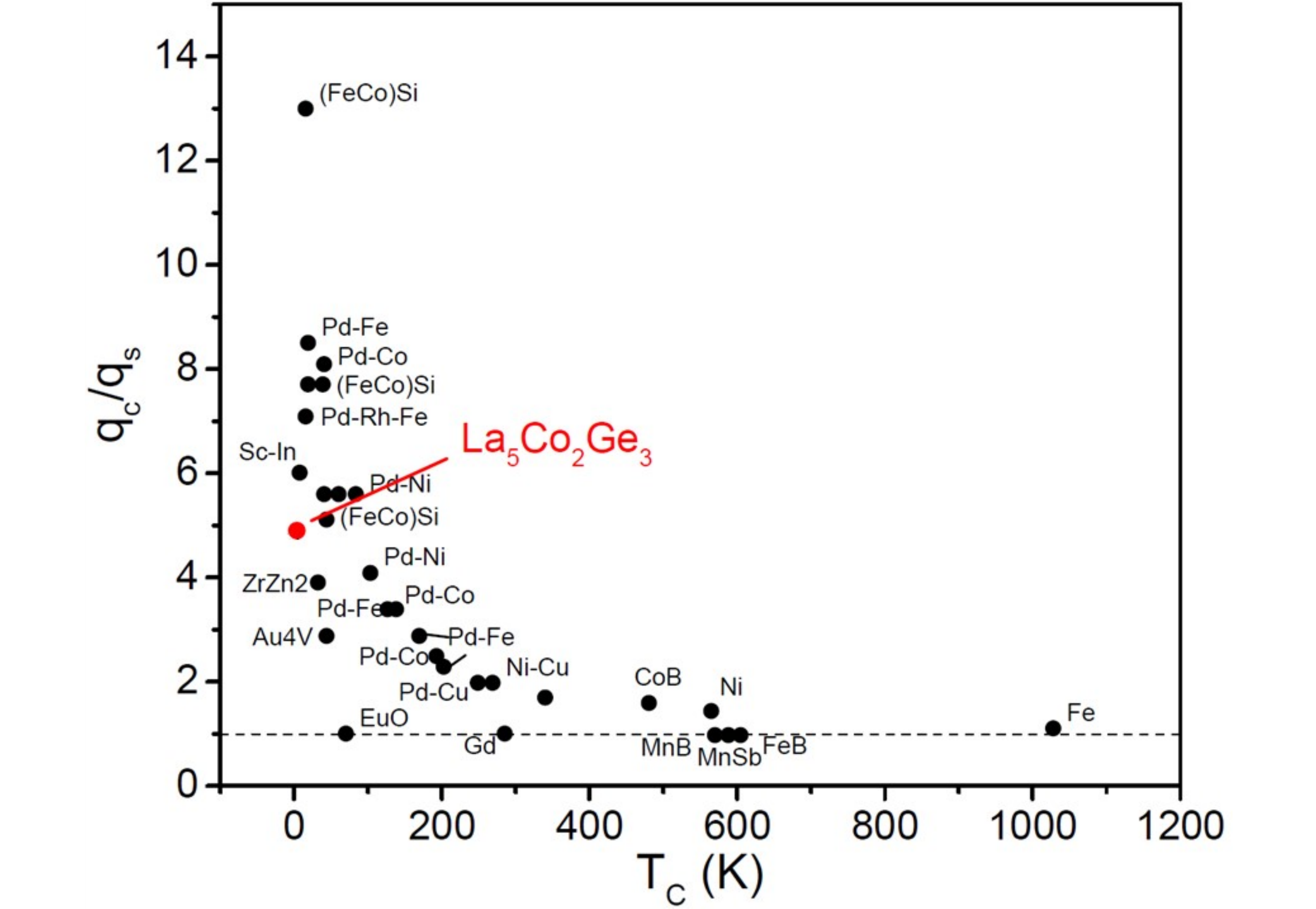}  
\caption{Rhodes-Wohlfarth ratio $q_c$/$q_s$ versus Curie temperature, $T_\mathrm C$, for various materials~\cite{Matthias_1958, Santiago_2017}. La$_{5}$Co$_{2}$Ge$_{3}$ is shown in red, where $q_c$ was determined from the effective moment obtained by fitting data from Fig.~\ref{fig:mt_perco} and $q_s$ determined from the saturated moment obtained from Fig.~\ref{fig:La_MH_toShow}.}
\label{fig:Rhode_Wohlfarth}
\end{figure}

We can compare La$_{5}$Co$_{2}$Ge$_{3}$ to other itinerant magnetic systems as shown in Fig.~\ref{fig:Rhode_Wohlfarth}. The Rhode-Wohlfarth ratio can be thought of as a measure of the change in magnetic moment as you change temperature ($\mu_{eff}$ inferred from high temperature data, $\mu_{sat}$ inferred from low temperature data). For La$_{5}$Co$_{2}$Ge$_{3}$, $q_c/q_s = 4.9$. Figure \ref{fig:Rhode_Wohlfarth} shows La$_{5}$Co$_{2}$Ge$_{3}$ is an intriguing combination of an ordered, line compound and one of the lowest Curie temperatures for transition-metal based ferromagnetism.

The thermodynamic, transport, and microscopic data presented on La$_{5}$Co$_{2}$Ge$_{3}$ all suggest that below 3.8~K there is small moment, itinerant, ferromagnetic ordering. The low-temperature, linear specific heat coefficient, $\gamma$, is also consistent with this. $\gamma = 40$~mJ/mol-K$^2$ is a rather large value, even for a compound with ten atoms per formula unit. This value can be put into context well by comparing it to $\gamma$ values for the Y(Fe$_x$Co$_{1-x}$)$_2$Zn$_{20}$ series~\cite{Jia2007}. Although YFe$_2$Zn$_{20}$ does not order magnetically, it is exceptionally close to the Stoner limit and has a $\gamma = 50$~mJ/mol-K$^2$. YCo$_2$Zn$_{20}$, on the other hand, is far from this limit and has a $\gamma = 20$~mJ/mol-K$^2$ (yielding a fairly standard 1~mJ/mol-atomic-K$^2$ value). Using a similar 1~mJ/mol-atomic-K$^2$ value for generic broad-band background, La$_{5}$Co$_{2}$Ge$_{3}$ has roughly 15~mJ/mol-Co-K$^2$, similar to the value found for Fe in YFe$_2$Zn$_{20}$. Comparison can also be made to LuFe$_2$Ge$_2$~\cite{Avila2004, Fujiwara2007} which has a spin-density wave type of itinerant antiferromagnetic ordering near 9~K and a $\gamma$-value of roughly 65~mJ/mol-K$^2$. 

Taken together, our data indicate that La$_{5}$Co$_{2}$Ge$_{3}$ is an ordered compound at the limit of low T$_C$ and high $q_c/q_s$. As such, it offers a chance to study how much further T$_C$ can be pushed, or tuned toward $T$=0, either by pressure or substitution before the anticipated avoided quantum criticality that is associated with metallic ferromagnetic systems is encountered~\cite{LaCrGe3Taufour2016, Kaluarachchi2018, CeTiGe3Kaluarachchi2018, Brando2016quantumferro}.

\section*{Acknowledgements}
We would like to thank A. Kreyssig for useful discussions and R. A. Ribeiro for assistance with magnetization measurements. This work is supported by the US DOE, Basic Energy Sciences under Contract No. DE-AC02-07CH11358 and the Gordon and Betty Moore Foundation’s EPiQS Initiative through Grant GBMF4411.
\\
\\
*T. Kong Current Address: Department of Physics, University of Arizona, Tucson, AZ 85721, USA (e-mail:tkong@email.arizona.edu)

\bibliographystyle{apsrev4-1}
%

\clearpage



\section*{Supplementary material}

# Supplementary Information: Exceedingly Small Moment Itinerant Ferromagnetism of Single Crystalline La$_5$Co$_2$Ge$_3$

S. M. Saunders,[1,2] L. Xiang,[1,2] R. Khasanov,[3] T. Kong,[1,2] Q. Lin,[1,4] S. L. Bud'ko,[1,2] and P. C. Canfield[1,2]

[1]*Ames Laboratory, U.S. DOE, Iowa State University, Ames, Iowa 50011, USA*
[2]*Department of Physics and Astronomy, Iowa State University, Ames, Iowa 50011, USA*
[3]*Laboratory for Muon Spin Spectroscopy, Paul Scherrer Institute, 5232 Villigen, Switzerland*
[4]*Department of Chemistry, Iowa State University, Ames, Iowa 50011, USA*

## SUPPLEMENTAL INFORMATION

### A. Structure Refinement

Single crystal X-ray diffraction intensity data for La$_5$Co$_2$Ge$_3$ were collected at room temperature using a Bruker SMART APEX II diffractometer (Mo K$\alpha$ radiation, $\lambda = 0.71073$ Å). Data reduction, integration, unit cell refinements, and absorption corrections were done with the aid of subprograms in APEX2[1,2]. Space group determination, Fourier Synthesis, and full-matrix least-squares refinements on F2 were carried out by in SHELXTL 6.1[3]. The direct methods in space group C2/m yielded a structural model containing five La, two Co, and three Ge independent sites. Separate refinements on occupancy parameters for Co and Ge sites revealed no partial occupancy and no Co/Ge mixing in this structure. Table S1 gives the crystal data and structure refinement for La$_5$Co$_2$Ge$_3$ and Table S2 lists the refined atomic positions and equivalent isotropic displacement parameters.

TABLE S1. Crystal data and structure refinement for La$_5$Co$_2$Ge$_3$.

| | |
|---|---|
| Empirical Formula | La$_5$Co$_2$Ge$_3$ |
| Formula Weight | 1030.2 g/mol |
| Space Group, Z | C2/m, 4 |
| Unit Cell Dimensions | a = 18.354(4) Å |
| | b = 4.3479(9) Å |
| | c = 13.279(3) Å |
| | $\beta$ = 109.592(2) |
| Z | 4 |
| Density (calculated) | 6.663 g/cm$^3$ |
| Reflections Collected | 30592[Rint = 0.0384] |
| Data / restraints / parameters | 5893 / 0 / 62 |
| Goodness-of-fit on F$^2$ | 1.179 |
| Final R indices [I>2$\sigma$(I)] | R1 = 0.0342, wR2 = 0.0530 |
| R indices (all data) | R1 = 0.0512, wR2 = 0.0564 |
| Largest diff. peak and hole | 1.316 and -1.486 e.$^-$3 |

The structure of La$_5$Co$_2$Ge$_3$ is part of a R$_5$Co$_2$Ge$_3$ family and represents a new structural type, with Pearson symbol of mS40. This structure type has been previously reported by Lin et. al[4]. The y coordinates for all

TABLE S2. The refined atomic positions and equivalent isotropic displacement parameters for La$_5$Co$_2$Ge$_3$

| Atom | Wyck. | Symm. | x | y | z | U$_{eq}$ (Å$^2$) |
|---|---|---|---|---|---|---|
| La1 | 4i | m | 0.0013(1) | 0 | 0.1371(1) | 0.0015(1) |
| La2 | 4i | m | 0.3171(1) | 0 | 0.4397(1) | 0.0015(1) |
| La3 | 4i | m | 0.3228(1) | 0 | 0.1090(1) | 0.0014(1) |
| La4 | 4i | m | 0.4998(1) | 0 | 0.3616(1) | 0.0015(1) |
| La5 | 4i | m | 0.6819(1) | 0 | 0.2300(1) | 0.0014(1) |
| Co1 | 4i | m | 0.0670(1) | 0 | 0.5212(1) | 0.0017(1) |
| Co2 | 4i | m | 0.5661(1) | 0 | 0.0240(1) | 0.0019(1) |
| Ge1 | 4i | m | 0.1325(1) | 0 | 0.7121(1) | 0.0015(1) |
| Ge2 | 4i | m | 0.1323(1) | 0 | 0.3749(1) | 0.0015(1) |
| Ge3 | 4i | m | 0.1466(1) | 0 | 0.0493(1) | 0.0015(1) |

atoms in this structure equal to zero, meaning that atoms in this structure are located either on planes at y = 0 or y = 1/2 arising from the C-center in space group C2/m. The structure appears as the ethylene-like Co$_2$Ge$_4$ fragments and the polyacene-like ribbons immersed in a sea of the rare-earth ions, c.f. Fig. S1. In this structure, Co-Co and Co-Ge bonds show strongest covalent bonding interactions, as indicated by respective bond distances (dCo-Co = 2.325-2.358 , dCo-Ge = 2.494-2.558 ). Notably, the separations for La1-La4, La2-La4, and La4-La4 pairs are smaller than the sum of Paulings metallic radii (3.648)[5], suggesting considerable covalent interactions among them. These pairs form two-dimensional honeycomb nets parallel to the bc plane, hexagons in the net are perpendicularly penetrated by Co-Co bonds (Figure 1). Sandwiched by the forgoing honeycomb nets, La2, La3 and La5 atoms form slabs of edge-sharing tetrahedra with slightly longer La-La distances (3.647-3.924 ).

### B. Magnetization Measurement and Analysis on Powder Sample

In order to better measure the high-temperature Curie-Weiss behavior, a powder sample was made by grinding 95 mg of single crystals and was measured in a field of 64 kOe as a function of temperature. The tempeartuer dependent $M(H)$ data (Fig.S2) manifest a clear Curie-Weiss-like behavior for 10 K$\leq T \leq$ 300 K. When the data are fit to Eqn. **??** with Curie Constant C= $N(\mu_{eff}\mu_B)^2/3k_B$, values of $\mu_{eff} = (1.10\pm0.05)$ $\mu_B/Co$,

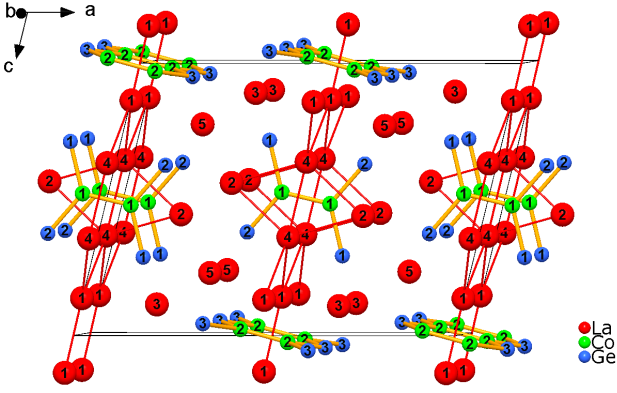

FIG. S1. Crystal structure of $La_5Co_2Ge_3$.

$\Theta = (-10.7 \pm 0.2)$ K, and $\chi_0 = 0.0038 \pm 0.0002$ were found.

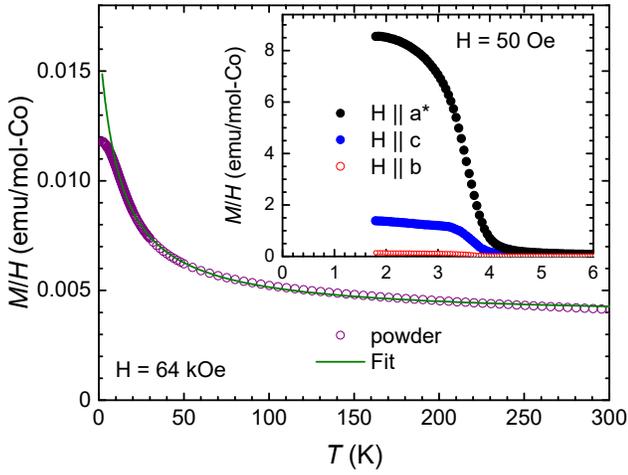

FIG. S2. Magnetic susceptibility of $La_5Co_2Ge_3$ powder measured at $H = 64$ kOe, with green line indicating fit of data (inset: low-temperature zoom of anisotropic magnetic susceptibility measured on single crystal at $H = 50$ Oe).

### C. ZF $\mu$SR Data Analysis Procedure

The time evolution of the muon-spin polarization, $P(t)$, was described by assuming the presence two internal fields $B_{int,1}$ and $B_{int,2}$ with the corresponding weight $f$ and $(1-f)$, respectively. In a case of $La_5Co_2Ge_3$ the presence of two internal fields most probably corresponds to the two muon-stopping sites. Note that the presence of multiple muon sites is often observed(see e.g. Refs. 6–8 and references therein). The following functional form was used:

$$P(t) = \frac{1}{3} e^{-\lambda_L t} + \frac{2}{3} \left[ f \cdot e^{-\lambda_{T,1} t} \cos(\gamma_\mu B_{int,1} t) \right. \\ \left. + (1-f) \cdot e^{-\lambda_{T,2} t} \cos(\gamma_\mu B_{int,2} t) \right]. \quad (1)$$

Here $\gamma_\mu = 2\pi\, 135.5$ MHz/T is the muon gyromagnetic ratio, and $\lambda_T$ and $\lambda_L$ are the transverse and the longitudinal exponential relaxation rates, respectively. The occurrence of 2/3 oscillating and 1/3 non oscillating $\mu$SR signal fractions originates from the spatial averaging in powder samples, where 2/3 of the magnetic field components are perpendicular to the muon-spin and cause a precession, while the 1/3 longitudinal field components do not.

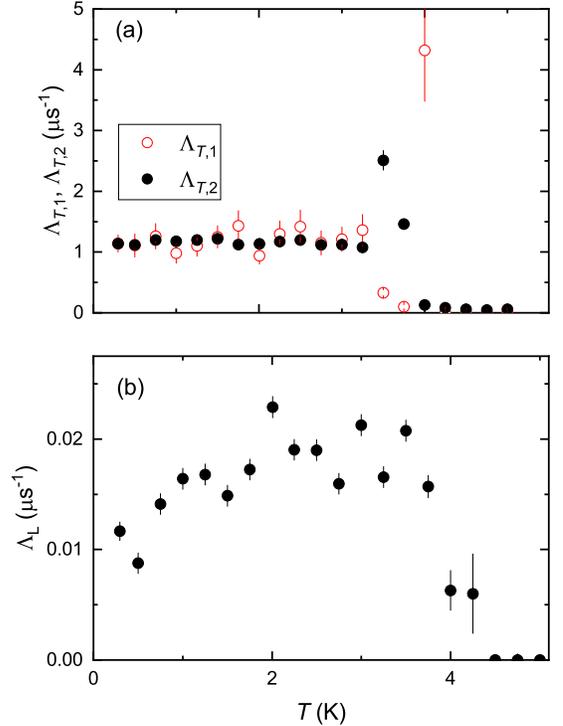

FIG. S3. (a) Temperature evolutions of the transversal relaxation rates $\lambda_{T,1}$ and $\lambda_{T,2}$. (b) Temperature evolution of the longitudinal relaxation rate $\lambda_L$.

The weight of the high-field component ($f$) is field independent and was found to be $f \simeq 0.19$. Temperature evolutions of the transversal ($\lambda_{T,1}$ and $\lambda_{T,2}$) and longitudinal relaxation ($\lambda_L$) are presented in Fig. S3. The similar temperature dependencies of $B_{int,1}$ and $B_{int,2}$ (see Fig. 7 in the main text) as well as $\lambda_{T,1}$ and $\lambda_{T,2}$ (see Fig. S3) confirms the presence of two muon-stopping sites in $La_5Co_2Ge_3$.



\end{document}